\documentclass[11pt]{article}
\usepackage[dvips]{graphicx}

\begin{document}

\begin{center}
\bf{GENERAL EXPRESSION FOR THE DIELECTRONIC RECOMBINATION
CROSS SECTION OF POLARIZED IONS WITH POLARIZED ELECTRONS} 
\end{center}

\vspace{5mm}
\begin{center}
A. Kupliauskien\.{e}\footnote{Corresponding author.
Tel.: +37052612723.\\E-mail address: 
akupl@itpa.lt. (A. Kupliauskien\.{e}).}, 
V. Tutlys
\end{center}

\begin{center}
{\small\em
Vilnius University Institute of Theoretical Physics 
       and Astronomy
A.Go\v{s}tauto 12, LT-01108 Vilnius, Lithuania }
\end{center}

\vspace{10mm}
{\bf Abstract}

A general expression for the differential cross section of
dielectronic recombination (DR) of polarized electrons and polarized
ions is derived by using usual atomic theory methods and is 
represented in the form of multiple expansions over spherical
tensors.
The ways of the application of the general expressions suitable 
for the specific
experimental conditions are outlined by deriving asymmetry parameters 
of angular distribution of DR radiation in the case of nonpolarized
and polarized ions and electrons.

\vspace{5mm}
\noindent
{\em PACS}: 34.80.Lx; 32.50.+d; 32.80.Hd\\
{\em Keywords}: Electron-ion recombination; Fluorescence; Inner-shell
                       excitation

\vspace*{10mm}
\noindent
{\bf 1. Introduction}

\vspace{5mm}
Dielectronic recombination (DR) and electron impact excitation are
the basic excitation mechanisms of x-ray production from high temperature
plasmas.
For the case of non-equilibrium, anisotropic plasmas, the polarization
and angular distribution of radiation can be useful for the investigation
of the properties of electron distribution function and plasma diagnostics 
\cite{Kazantsev}.
In tokamak an other laboratory devices, the highly charged ions can 
be aligned \cite{Kazantsev,Inal}, 
therefore the expression for DR cross section describing
the polarization state of ions and electrons in the initial state and
radiation in the final state are of importance.
The alignment of the doubly excited state, i.e. a non-uniform
occupation of magnetic sublevels, causes the anisotropic emission 
of a photon relative to the direction of an incident ion beam 
\cite{Zakowicz}.
The values of calculated DR cross section for  various directions can
change up to several tens of percent when the anisotropy of DR
radiation is taken into account \cite{Balashov1994}.

Density matrix formalism \cite{Balashov2000} is the usual method
for the derivation of the expressions for the differential cross
sections of DR of polarized ions with polarized electrons
\cite{Balashov1994,Balashov2000}.
Recently in the case of fully relativistic treatment, the expression
for the differential cross section of DR was derived \cite{Zakowicz}
by using a projection operator formalism \cite{LaGatutta}.
The aim of the present work is the derivation of general expression
in nonrelativistic approximation for the differential cross section of 
DR process by applying alternative method based on the atomic theory
methods \cite{Jucys1965,Merkelis,KRT2000,KRT2001}.
The polarization state of all particles in both the initial and final
states are described in this approximation.
The practical applications of the general expression for the specific
experimental conditions are outlined.
The expression for the asymmetry parameter of the angular distribution
of radiation in DR of non-polarized and aligned ions with non-polarized
and polarized electrons demonstrates the way of the derivation of more
simple expressions.

\vspace*{ 5mm}
\noindent
{\bf 2.  General expression}

\vspace{5mm}
The process of DR can be written as follows
\begin{equation}
A^+(\alpha_0 J_0M_0)+e^-({\rm\bf p} m) \to
A^{**}(\alpha_1 J_1) \to \Biggl\{
\begin{array}{l}  A(\alpha_2 J_2M_2) + 
h\nu(\epsilon_{q},{\rm\bf k}_{0}),\\
\\
 A^+(\alpha_3 J_3M_3) + e^-({\rm\bf p}_1m_1).
\end{array}
\end{equation}

\noindent
It is an example of two-step process.
The first step is resonant electron capture that is inverse process to
 Auger decay.
The next step is radiative or Auger decay
those expressions for the differential probability were obtained
\cite{KT2003,KT2004} in the case of orientated and aligned ions
following photoionization of atoms.
Two-step approximation for DR may be applied if the interference
 with the radiative recombination
is neglected and the summation over intermediate states $J_1M_1$,
that usually occurs 
in second-order perturbation theory, is limited to a single resonance.
Then, only a summation over the magnetic substates that are not registered
is retained.
DR process is finished when the photon is emitted. 

In two-step approximation, the cross section for DR may be written 
 as:

$$
\frac{d\sigma(\alpha_0 J_0M_0{\rm\bf p}m \to \alpha_1 J_1 \to
\alpha_2 J_2M_2 \epsilon_{q}{\rm\bf k}_{0})}{d\Omega}= 
\frac{\rho_f}{\rho_i}
\sum_{M_1,M'_1}
\langle \alpha_2 J_2M_2 \epsilon_{q}{\rm\bf k}_{0}|H'|
\alpha_1 J_1M_1\rangle
$$
$$
\times
\langle \alpha_2 J_2M_2 \epsilon_{q},{\rm\bf k}_{0}|H'|
\alpha_1 J_1M'_1\rangle^*
\langle \alpha_1 J_1M_1|H^e|(\alpha_0 J_0M_0{\rm\bf p} m\rangle
\langle \alpha_1 J_1M'_1|H^e|(\alpha_0 J_0M_0{\rm\bf p} m\rangle^*
$$
\begin{equation}
\times
[(E-E_1)^2 + \Gamma^2/4]^{-1}.
\end{equation}

\noindent
Here $H'$ and $H^e$ is the radiative decay and electrostatic interaction
operators, respectively, 
$d\Omega$ is the solid angle of the emission of radiation,
$E_1$ and $E$ is the energy of the intermediate and initial state of the system
atom+electron, respectively, and
$\Gamma$ denotes the decay width of the intermediate state that includes
both radiative and nonradiative decay channels.
In (2), $\rho_i$ and $\rho_f$ denote the flux of incoming  electrons
and the density of final states, respectively.
Atomic system of units is used.

In two step approximation ($E\approx E_1$), the general expression for DR (2)
in the case of the interaction of polarized ion with polarized
electron may be obtained by applying the methods described in 
\cite{KRT2001,KT2003} and is as follows:

$$
\frac{d\sigma(\alpha_0 J_0M_0{\rm\bf p} m \to \alpha_1 J_1 \to
\alpha_2 J_2M_2 \epsilon_{q},{\rm\bf k}_{0})}{d\Omega}= 
 \frac{\rho_f}{\rho_i}
\sum_{K_1,N_1}
W^c_{K_1 N_1}(\alpha_0 J_0M_0{\rm\bf p} m \to  \alpha_1 J_1)
$$
\begin{equation}
\times
\frac{dW^r_{K_1 N_1}(\alpha_1 J_1 \to 
\alpha_2 J_2M_2 \epsilon_{q},{\rm\bf k}_{0})}{d\Omega}
[(E-E_1)^2+\Gamma^2/4]^{-1}.
\end{equation}

The resonant electron capture cross section $W^c$ is reversed to that of Auger
decay and is defined by \cite{KT2003}
$$
W^c_{K_1N_1}(\alpha_1 J_1 \to \alpha_0J_0 M_0{\bf p}m)
=
\sum_{K,K_0,K_\lambda,K_s}
{\cal A}^a(K_1,K_0,K_\lambda,K_s,K)
\sum_{N,N_0,N_\lambda,N_s}
\left[\begin{array}{ccc}
K_\lambda & K_s & K\\N_\lambda &N_s &N
\end{array} \right]
$$
\begin{equation}
\times
\left[\begin{array}{ccc}
K_0 & K & K_1\\N_0 &N &N_1
\end{array}\right]
T^{*K_0}_{N_0}(J_0,J_0,M_0|\hat{J}_0)\;
T^{*K_s}_{N_s}(s,s,m|\hat{s}) \;
\sqrt{4\pi} Y^*_{K_\lambda N_\lambda}(\theta_1,\phi_1).
\end{equation}

\noindent
Here
\begin{equation}
T^K_N(J,J',M|\hat{J})=
(-1)^{J'-M}\left[\frac{4\pi}{2J+1}\right]^{1/2}
\left[\begin{array}{ccc}
J & J' & K\\M & -M & 0\end{array}\right]
Y_{KN}(\hat{J}),
\end{equation}
$$
{\cal A}^a(K_1,K_0,K_\lambda,K_s,K)=
2\pi\sum_{\lambda_1,j_1,\lambda_2,j_2}
\langle \alpha_0 J_0\varepsilon\lambda_1(j_1)J_1||H||\alpha_1J_1\rangle
\langle \alpha_0 J_0\varepsilon\lambda_2(j_2)J_1||H||\alpha_1J_1\rangle^*
$$
\begin{equation}
\times
 (2J_1+1)
\left[(2\lambda_1+1) (2\lambda_2+1)
(2j_1+1)(2j_2+1)(2J_0+1)(2s+1)(2K+1)\right]^{1/2}
$$
$$
\times
         \left\{
\begin{array}{ccc}
J_0 &j_1 & J_1\\ J_0 & j_2 & J_1\\ K_0 & K & K_1
\end{array}
          \right\}
         \left\{
\begin{array}{ccc}
\lambda_2 &s & j_2\\ K_\lambda & K_s & K\\ \lambda_1 & s & j_1
\end{array} \right\}
(-1)^{\lambda_2}
\left[\begin{array}{ccc}
\lambda_1 &\lambda_2 &K_{\lambda}\\ 0 & 0 & 0
\end{array}\right].
\end{equation}

The expression for the radiative decay probability $dW^r/d\Omega$ is 
\cite{KT2004}:
$$
\frac{dW^r_{K_1N_1}(\alpha_1J_1 \to
\alpha_2J_2M_2 \hat{\epsilon}_{q}{\rm\bf k}_{0}) } {d\Omega} =
\sum_{K_r,K_2,k,k'}
{\cal A}^r(K_1,K_r,K_2,k,k')
$$
\begin{equation}
\times
\sum_{N_r,N_2}
\left[\begin{array}{ccc}
K_1 & K_r & K_2\\N_1 &N_r &N_2
\end{array}\right]
T^{K_2}_{N_2}(J_2,J_2,M_2|\hat{J}_2)\;
T^{*K_r}_{N_r}(k,k',q|\hat{{\rm\bf k}}_{0}),
\end{equation}
$$
{\cal A}^r(K_1,K_r,K_2,k,k') =
C(k,k')
( \alpha_2J_2 ||Q^{(k)}||\alpha_1J_1)
( \alpha_2J_2 ||Q^{(k')}||\alpha_1J_1)^*
$$
\begin{equation}
\times
\left[\frac{(2K_1+1)(2J_2+1)(2k+1)}{2K_2+1}\right]^{1/2}
\left\{\begin{array}{ccc}
J_1 & K_1 &J_1 \\ k & K_r & k' \\ J_2 & K_2 & J_2
\end{array}\right\} .
\end{equation}
In (8), the relation
\begin{equation}
\langle \alpha_2J_2||Q^{(k)}||\alpha_1J_1\rangle =
(2J_2+1)^{-1/2} (\alpha_2J_2||Q^{(k)}||\alpha_1J_1)
\end{equation}
\noindent
is applied.

Below the general expressions (4), (6), (7) and (8) are used 
to obtain some
special expressions for specific experimental conditions.

\vspace*{ 5mm}
\noindent
{\bf 3. Special cases}

\vspace{5mm}
The tensor (5) describes the orientation of the angular momentum
with respect to the laboratory $z$ axis.
In the case of the magnetic components are not registered, the
summation over them of (5) leads to $\delta(K,0)\delta(N,0)$.
Thus, to obtain the expression of DR cross section describing
angular distribution of radiation in the case of nonpolarized
ions and electrons, one needs to insert $K_0=N_0=K_s=N_s=K_2=N_2=0$
and $K_1=K_\lambda=K=K_r=$~even into (4). 
Taking into account the $z$ axis coinciding with the direction
of electrons ($N_1=0$) we can write DR cross section in well
known form \cite{Chen}:
\begin{equation}
\frac{d\sigma(\alpha_0 J_0 \to \alpha_1 J_1 \to
\alpha_2 J_2{\rm\bf k}_{0})}{d\Omega}= 
\frac{\sigma(\alpha_0 J_0 \to \alpha_1 J_1 \to\alpha_2 J_2)}
{4\pi}\left[1+\sum_{K_1>0}\beta_{K_1}P_{K_1}(\cos\theta)\right].
\end{equation}
Here $\sigma(\alpha_0 J_0 \to \alpha_1 J_1 \to\alpha_2 J_2)$ is
DR cross section, and
the asymmetry parameter of the angular distribution of DR
radiation (the angle $\theta$ is measured from the direction of
electrons) is defined by
\begin{equation}
\beta_{K_1}=\sum_{k,k'}(-1)^{k-q}
\sqrt{(2k+1)(2K_1+1)}
\left[\begin{array}{ccc}
k & k' & K_1\\q &-q &0\end{array} \right]
\frac{{\cal A}^a(K_1,0,K_1,0,K_1)}{{\cal A}^a(0,0,0,0,0)}
\frac{{\cal A}^r(K_1,K_1,0,k,k')}{{\cal A}^r(0,0,0,k,k')}.
\end{equation}
In the case of electrical dipole radiation, $k=k'=1$, and
the expression for $\beta_2$ coincides with (9) from \cite{Chen}.

To obtain an expression for the dependence of angular distribution
of DR radiation on the polarization state of ions we need to 
perform summation of (4) over $m=\pm 1/2$ states of the electron 
and $M_2$ of the final state of recombined ion that lead to
$K_s=N_s=K_2=N_2=0$, $K_\lambda=K=$~even, $K_1=K_r$. 
The choice of the laboratory $z$ axis along the direction of
electrons gives $N_\lambda=0$ because of 
$Y_{K_\lambda N_\lambda}(0,0)=[(2K_\lambda+1)/4\pi]^{1/2}
\delta(N_\lambda,0)$.
Then
$$
\frac{d\sigma(\alpha_0 J_0 M_0 \to \alpha_1 J_1 \to
\alpha_2 J_2{\rm\bf k}_{0})}{d\Omega}= 
\frac{\rho_f}{2\rho_i[(E-E_1)^2+\Gamma^2/4]}
\sum_{K_1,N_1,k,k'} {\cal A}^r(K_1,K_1,0,k,k')
T^{*K_1}_{-N_1}(k,k',q|{\rm\bf k}_0)
$$
\begin{equation}
\times
\sum_{K_0,K_\lambda}
{\cal A}^a(K_1,K_0,K_\lambda,0,K_\lambda)
\left[\begin{array}{ccc}
K_1 & K_0 & K_\lambda\\-N_1 &N_1 &0\end{array} \right]
T^{*K_0}_{N_1}(J_0,J_0,M_0|\hat{J}_0).
\end{equation}

In the case when the final state of an ion is not registered
and laboratory $z$ axis chosen along the direction of
incoming electrons, the expression for DR of nonpolarized
ions with polarized electrons can be obtained by averaging
with respect of magnetic states of ion and summation over $M_2$
of (4).
Then $K_0=N_0=K_2=N_2=0$, $K_1=K=K_r$, $N_\lambda=0$, and
$$
\frac{d\sigma(\alpha_0 J_0 m \to \alpha_1 J_1 \to
\alpha_2 J_2{\rm\bf k}_{0})}{d\Omega}= 
\frac{\rho_f}{(2J_0+1)\rho_i[(E-E_1)^2+\Gamma^2/4]}
\sum_{K_1,N_1,k,k'} {\cal A}^r(K_1,K_1,0,k,k')
$$
\begin{equation}
\times
T^{K_1}_{-N_1}(k,k',q|{\rm\bf k}_0)
\sum_{K_s,K_\lambda}
A^a(K_1,0,K_\lambda,K_s,K_1)
\left[\begin{array}{ccc}
K_s & K_1 & K_\lambda\\N_1 &-N_1 &0\end{array} \right]
T^{*K_s}_{N_1}(s,s,m|\hat{s}).
\end{equation}

The last sums in (12) and (13) can be called differential 
alignments similar to that introduced in \cite{Surzhydov}.

\vspace*{5mm}
\noindent
{\bf 4. Concluding remarks}

The general expression for  DR differential cross section of polarized 
ions and polarized electrons is obtained in two-step approximation.
It is presented in the form of the multipole expansions over the states 
of all particles
participating in the process that is very convenient for investigations, 
since the geometrical and dynamical parts are separated.
This form coincides with that of the expansion over state multipoles 
(statistical tensors) widely used in the density matrix formalism.
A simple way to derive the expressions for the special cases by using 
the general expression for the DR is described.
The expressions for the asymmetry parameter of the angular distribution of 
radiation is obtained in the case of DR of nonpolarized atoms with 
nonpolarized and polarized electrons.

\end{document}